\documentclass[namedreferences]{SolarPhysics}

\usepackage{graphicx}        
\usepackage{amssymb}        
\usepackage{url}             
\newcommand{\fb}{}
\newcommand{\SN}{S\!N}

\def\copyrightline{}

\begin{document}
\begin{article}
\begin{opening}
\newcommand\mytit{Addition to the Gnevyshev-Ohl rule and prediction of solar cycle 25}
\title{\mytit}
\author{Yu.~A.~\surname{Nagovitsyn}$^{1,2}$}
\author{V.~G.~\surname{Ivanov}$^{1}$}
\runningauthor{Nagovitsyn and Ivanov}
\runningtitle{\mytit}
\institute{$^{1}$Central Astronomical Observatory at Pulkovo, Saint-Petersburg, Russia\\
$^2$State University of Aerospace Instrumentation, St. Petersburg, Russia}

\begin{abstract}
In addition to the Gnevyshev-Ohl rule (GOR), the relation of the odd
cycle with the subsequent even one in the 22-year Hale solar cycle was
found. It is shown that 3~years before the 11-year minimum $m$, the value of the 
relative sunspot number $\SN$ in an odd cycle is closely related to the
value of the maximum in the next even cycle (correlation coefficient $\rho=0.94$),
and the same relation of an odd cycle with the previous even one is
weaker. Like GOR, cycles are linked in pairs, but opposite to the
Rule.

Based on this result, we propose to use $\SN_{m-3}$ on the descending phase of
the previous odd cycle as a precursor of the subsequent EVEN cycle
(Figure~\ref{fi3}a) --- a precursor called MI3E. For the prediction of an odd
cycle or a prediction without consideration of parity (as in the
article by \opencite{brajsa22}), this method gives less reliable
results.

To predict the amplitude of an ODD cycle, we propose to use the
precursor of the seventh year to its maximum $M$ MA7O --- $\SN_{M-7}$ on the
descending phase of the previous even cycle (Figure~\ref{fi3}b). It turned out
that in this case, we can also predict the years near the maximum with
a high correlation coefficient ($\rho=0.90{-}0.94$).

Thus, the proposed approaches allow us to predict cycles of different
parity. According to our prediction, the current solar Cycle 25 in
2023 will reach a maximum of 154 units with a prediction error of $\pm25$
(68\% confidence) and $\pm53$ (95\% confidence). In 2024, $\SN$ will be almost
as high as in 2023 --- 147 units, so with smaller time averaging scales,
the maximum will fall at the end of 2023.
\end{abstract}
\end{opening}

\section{Introduction}

As noted in (\opencite{nag09}), ``The empirical Gnevyshev-Ohl
rule (1948), below referred to as GOR or the Rule, is one of the most
puzzling properties of the solar cyclicity.'' \inlinecite{gnevohl48},
based on the consideration of a relative sunspot numbers $\SN$ from 1700
to 1944, found that the area under the curve (total power) of the
11-year cycle $\sum \SN$ has a high correlation for the pair of even --
subsequent odd cycles and weakly correlates for the pair of odd --
subsequent even cycles. Thus, they concluded that the 22-year Hale
cycle begins with an even cycle. This is a strange circumstance since
the resulting rule divides cycles into pairs, and neighboring pairs
are weakly related. One of the questions we raise in this article is:
in addition to the parameter $\sum \SN$, are there cycle parameters that
connect consecutive cycles in different ways, including in the odd --
subsequent even cycle pair?

In addition, we touch on a purely practical question of solar cycle
prediction: how will look the recently started Cycle 25, and what will
be its amplitude when the moment of maximum comes. As it turned out,
in the context under consideration, the prediction depends, and quite
critically, on the parity of cycles, i.e., the global organization of
the Sun's magnetic field in the light of the Hale cycle.

\section{Gnevyshev-Ohl rule}

\begin{figure}
\begin{center}
\fb{\includegraphics[width=0.45\textwidth]{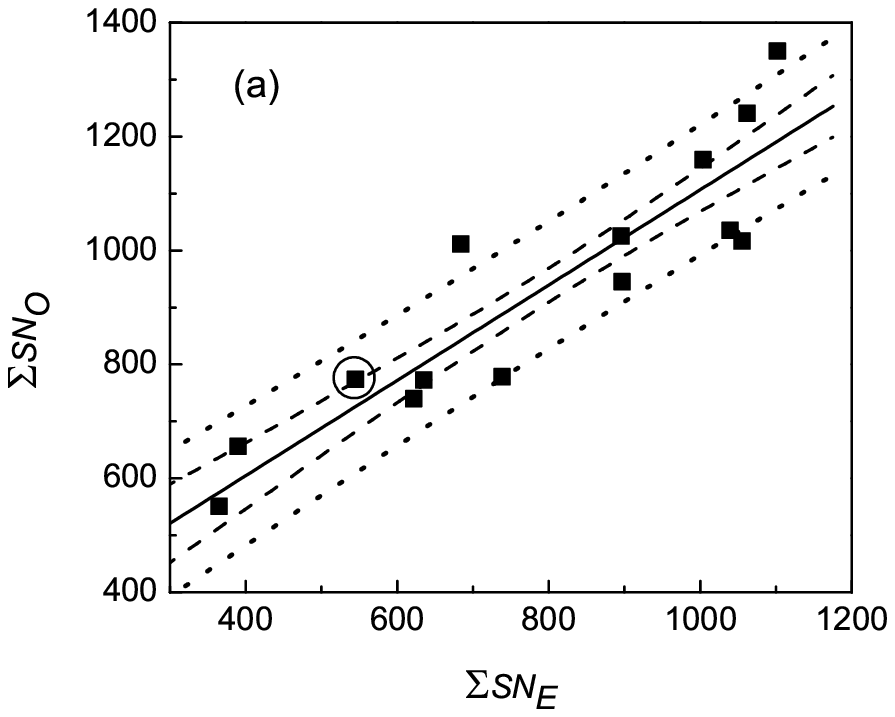}}
\fb{\includegraphics[width=0.45\textwidth]{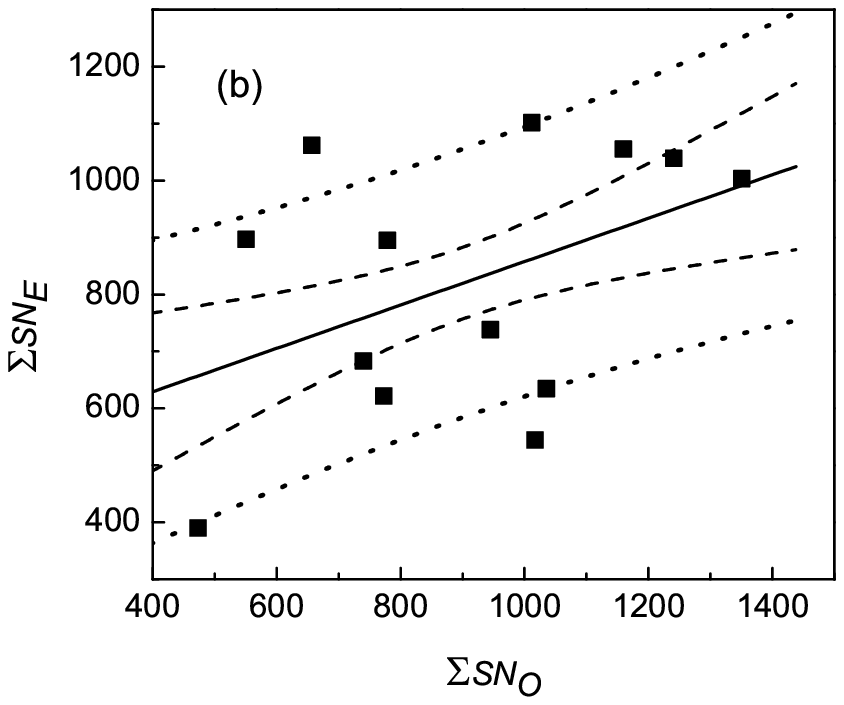}}
\end{center}
\caption{
Empirical relations (squares) and regressions (solid lines)
for two consecutive cycles: (a) $\sum \SN_O = (270\pm99)+(0.84\pm0.12)\cdot\sum \SN_E$, $\rho=0.906$;
(b) $\sum \SN_E = (480\pm220)+(0.38\pm0.24)\cdot\sum \SN_O$, $\rho=0.438$, where $\rho$ is the Pearson
correlation coefficient. Standard deviations are dashed lines, and the
prediction intervals (with 68\% confidence) are dotted lines. The
circled point, which was not used in the regressions' calculation,
corresponds to the prediction of the Cycle 25 (see Figure~\ref{fi6}b).
}
\label{fi1}
\end{figure}

Almost three-quarters of a century has passed since the publishing of
the article by Gnevyshev and Ohl. Since that time, new 11-year cycles
appeared, adding new statistical data, and the version of $\SN$ used by
these authors was replaced by a version called $\SN$ 2.0 (\opencite{clette14}, \citeyear{clette16}).
Therefore, using SILSO data \url{https://www.sidc.be/silso/}
let's plot the dependencies $\sum \SN_E = f_1(\sum \SN_O)$,
$\sum \SN_O = f_2(\sum \SN_E)$, and calculate the correlation
coefficients $\rho$ for them according to modern material, starting from
the year 1700. The ``E'' and ``O'' subscripts hereinafter indicate even
and odd cycles, respectively. As in the original work of \inlinecite{gnevohl48},
we exclude from consideration a pair of Cycles 4 and 5.
Figure~\ref{fi1} illustrates the results. The confidence interval $\pm1\sigma$ is
selected (the probability of falling within is 68\%).
 
There is an evident confirmation of the GOR. Thus, this rule captures
the causation of the odd cycle with the previous even one for the
modern version of $\SN$ 2.0 as well (Figure~\ref{fi1}a), while the even cycle
within the Rule is not related to the previous odd one (Figure~\ref{fi1}b).

A summary of possible methods for predicting the amplitudes of 11-year
cycles can be found in the review by \inlinecite{petrovay20}. In particular,
it contains the information about precursor methods.

The precursor methods are based on the idea that the cycle begins
before the sunspot solar cycle minimum between the old and the new
cycle. This idea does not contradict the general approach of the solar
dynamo and can be used to predict the amplitude of the future maximum
in the epoch near the previous minimum. A.I. Ohl (\citeyear{ohl66}) was one of the
first to propose such a method: as a precursor, he used geomagnetic
activity. The geomagnetic activity has often been used for solar
activity cycle prediction (for example, \opencite{pesnell14}). \inlinecite{sval05}
used a polar field. \inlinecite{wilson98} proposed a comprehensive prediction for a number of
cycle parameters. ``Older'' and less-known prediction methods, including
those with the approach of precursors, are contained in Vitinsky's
monograph (\citeyear{vit73}).

According to \inlinecite{macint20}, the international group of
experts Solar Cycle 25 Prediction Panel, (SC25PP) concluded that the
sunspot solar Cycle 25 will be similar in amplitude to Cycle 24; the
maximum will occur no earlier than 2023 and no later than 2026 with
the sunspot number from 95 to 130. However, at the beginning of this
year, Leamon and McIntosh announced that Cycle 24 had finally ended
(Termination Event occurred), and this allows us to look at the
prediction of \inlinecite{macint20} differently:
\url{https://spaceweatherarchive.com/2022/02/25/the-termination-event-has-arrived}.
According to the new prediction, the maximum of the Cycle 25 will be
about 190 (140--240), i.e., close in amplitude to the rather large
Cycle 23.

Most recently, an article by \inlinecite{brajsa22} was published, in
which the maximum of the started Cycle 25 is predicted as $121 \pm 33$ ---
i.e., low.

Thus, at the moment there is uncertainty about the future amplitude of
the Cycle 25, and the scatter of predictions is large.

\section{Cycle precursor associated with the minimum phase}

In the article by \inlinecite{brajsa22}, it is proposed to choose
values of $\SN$ three years before the minimum of the cycle as a
precursor of the maximum amplitude of the cycle. The maximum smoothed
monthly averages were used. We will use annual averages in our work ---
this is not a fundamental difference for the prediction. The interval
for the study is the same as that of \inlinecite{brajsa22} --- from the
year 1749 to the present.

Let's make some remarks. According to SILSO data, Cycle 20 has a
fairly flat maximum with maximum amplitude in the year 1968. At the
same time, $\SN$ show a different maximum according to Kislovodsk data
\url{http://www.solarstation.ru/archiv}, the sunspot group number $G\!N$
according to \inlinecite{svalshat16}, \inlinecite{usos16} data, and the sunspot area according to Greenwich data
\url{https://solarscience.msfc.nasa.gov/greenwch.shtml}: in 1970 --- see Table~\ref{tab1}, the values in the maxima years are underlined. Based on this, in
our work, we will assume that the maximum of the 20th cycle
corresponds to the middle of the year 1970--1970.5.

\begin{table}
\caption{
Average annual values of various indices of solar activity in Cycle 20.
}\label{tab1}
\begin{tabular}{cccccc}    
\hline
Year & $\SN$, & $\SN$, & $G\!N$, & $G\!N$, & Area, \\
& SILSO & Kislovodsk & Svalgaard & Usoskin &  $\mu$sh, \\
& & & \& Schatten, 2016 & el al., 2016 & Greenwich \\
\hline
1964.5 & 15.0  & 12.9  & 0.88 & 0.845 & 54    \\
1965.5 & 22.0  & 20.4  & 1.21 & 1.203 & 113   \\
1966.5 & 66.8  & 61.2  & 3.32 & 3.635 & 593   \\
1967.5 & 132.9 & 133.3 & 7.07 & 7.934 & 1519  \\
1968.5 & \underline{150.0} & 142.2 & 7.14 & 8.129 & 1570  \\
1969.5 & 149.4 & 139.3 & 7.27 & 7.947 & 1450  \\
1970.5 & 148.0 & \underline{148.3} & \underline{8.04} & \underline{8.968} & \underline{1601}  \\
1971.5 & 94.4  & 109.6 & 5.49 & 6.081 & 990   \\
1972.5 & 97.6  & 108.8 & 5.33 & 5.955 & 917   \\
1973.5 & 54.1  & 54.0  & 2.95 & 3.235 & 458   \\
1974.5 & 49.2  & 46.7  & 2.69 & 2.824 & 399   \\
1975.5 & 22.5  & 19.9  & 1.19 & 1.252 & 166   \\
1976.5 & 18.4  & 16.7  & 1.08 & 1.122 &170    \\
\hline
\end{tabular}
\end{table}

\begin{figure}
\begin{center}
\fb{\includegraphics[width=0.45\textwidth]{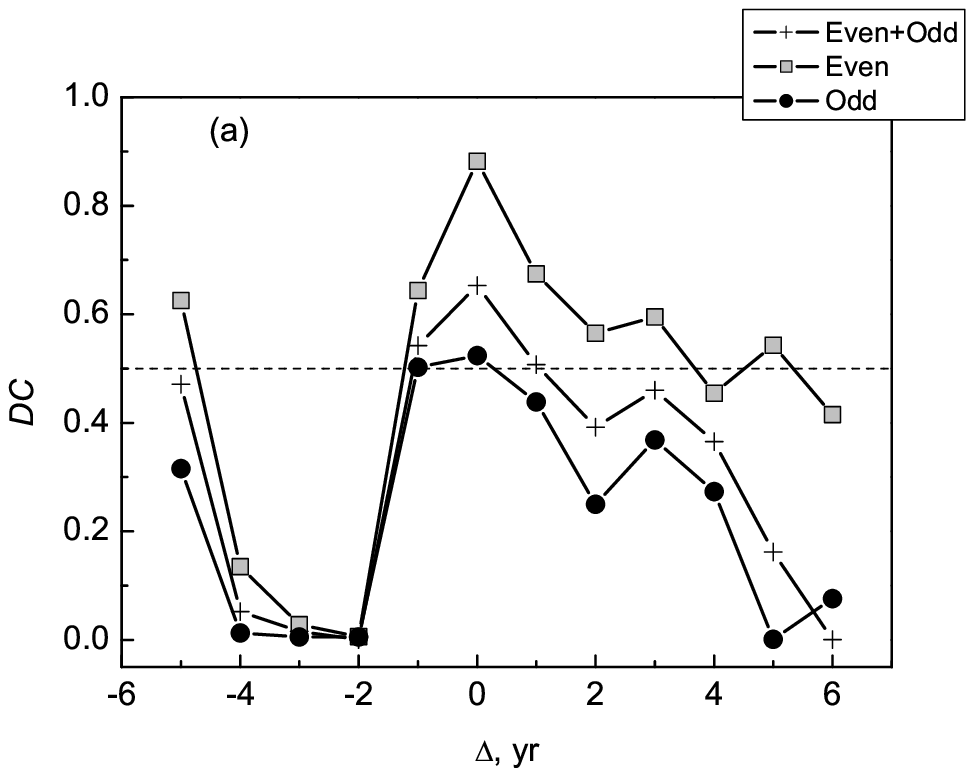}}
\fb{\includegraphics[width=0.45\textwidth]{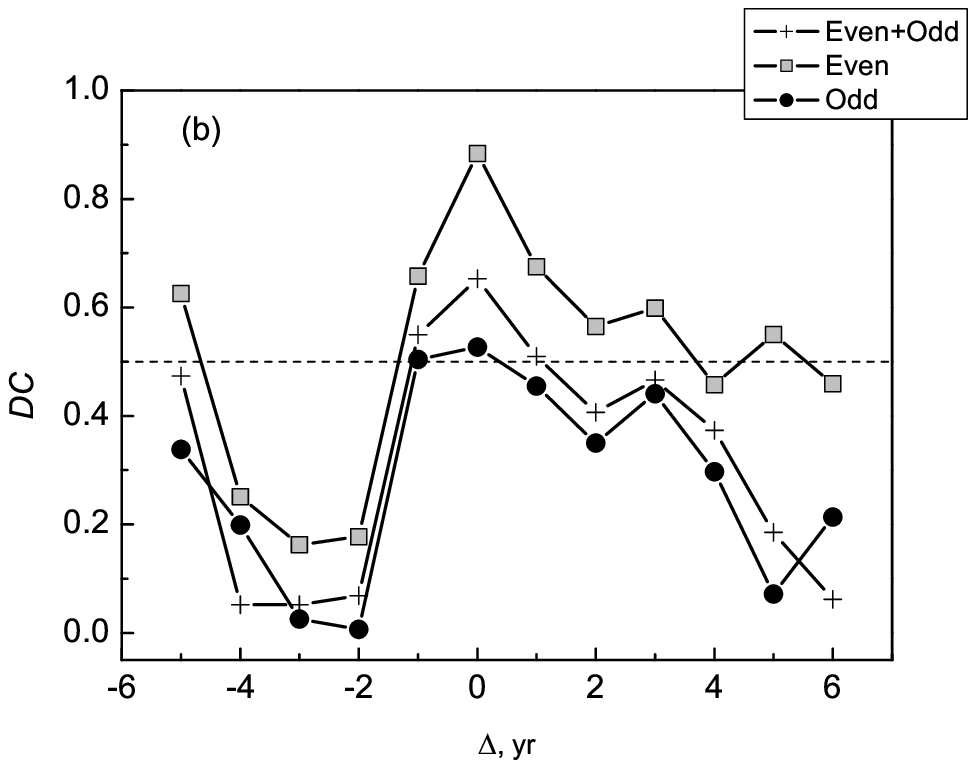}}
\end{center}
\caption{
Coefficients of the determination for relations $\SN_{m-3}$ with $\SN_{M+\Delta}$ (a)
for the linear form~(\ref{eq1}), (b) for the nonlinear form~(\ref{eq2}). The parity of
the cycles and their mixture corresponds to the legend.
}
\label{fi2}
\end{figure}

Let's take the values of $\SN$ three years before the minimum and
calculate their correlations with the values of different years that
are separated by $\Delta$ years from the next maximum. Negative values of $\Delta$
will correspond to the years before the maximum, positive --- after.
Zero is the maximum itself. The minimum will be denoted by $m$, the
maximum by $M$. The parity of the cycle will be determined by the
predicted maximum. We will build linear relations of the following
form using the least squares method (LSM):
\begin{equation}
\SN_{M+\Delta} = a + b\,\SN_{m-3}\,,
\label{eq1}
\end{equation}
and weakly non-linear ones:
\begin{equation}
\SN_{M+\Delta} = a + b\,\SN_{m-3} + c\,\SN_{m-3}^2\,,
\label{eq2}
\end{equation}
and also calculate the coefficients of determination for them $DC\equiv\rho^2$. This
coefficient is preferable because it has a clear meaning: it is the
proportion of dispersion that can be described using the appropriate
regression. The results are shown in Figure~\ref{fi2} separately for even,
odd, and all cycles together. The dashed line shows the upper limit of
the ``acceptable'' level of correlation for the prediction from our
point of view: $DC=0.5$ ($\rho=0.7$).

The first thing that can be seen in Figure~\ref{fi2}: taking into account the
nonlinearity does not lead to a noticeable improvement in
correlations. But that's not the main thing. It turns out that the
tightness of the relation $\SN_{m-3}$ with $\SN$ in the maximum and the years
closest to it strongly depends on the parity of the predicted cycle.
We can predict an even cycle with a coefficient of determination D =
0.882 (respectively, with $\rho = 0.94$), while an odd cycle --- only with $DC
= 0.524$ ($\rho = 0.72$).

According to the prediction in the style of the article by \inlinecite{brajsa22},
i.e., if we do not take into account the parity of cycles,
we get for a maximum of Cycle 25: $\SN_M = 129\pm34$, $\rho=0.808$, which is quite consistent with
the values obtained in their article. According to the nonlinear
prediction~(\ref{eq2}), our same values were obtained --- taking into account
the possible nonlinearity does not give anything additional: $\SN_M = (71\pm17) + (1.45\pm0.23)\SN_{m-3}$. It is
also close to the received by \inlinecite{brajsa22}.

Now let's see what the regressions and the prediction will look like
if we take into account the parity of the cycles. For even cycles
regression
\begin{equation}
\SN_M = (50\pm14) + (1.60\pm0.18) \SN_{m-3}\,, \quad \rho = 0.939\,,
\label{eq3}
\end{equation}
If our predicted cycle were even, then its maximum would be $\SN_M=113\pm21$.
Confidence intervals due to the high correlation coefficient are very
small. However, Cycle 25 is odd. We calculate the regression for odd
cycles:
\begin{equation}
\SN_M = (91\pm29) + (1.32\pm0.40) \SN_{m-3}\,, \quad \rho = 0.724
\label{eq4}
\end{equation}
and the prediction itself:
\begin{equation}
\SN_M = 144\pm44\,.
\label{eq5}
\end{equation}

\begin{figure}
\begin{center}
\fb{\includegraphics[width=0.45\textwidth]{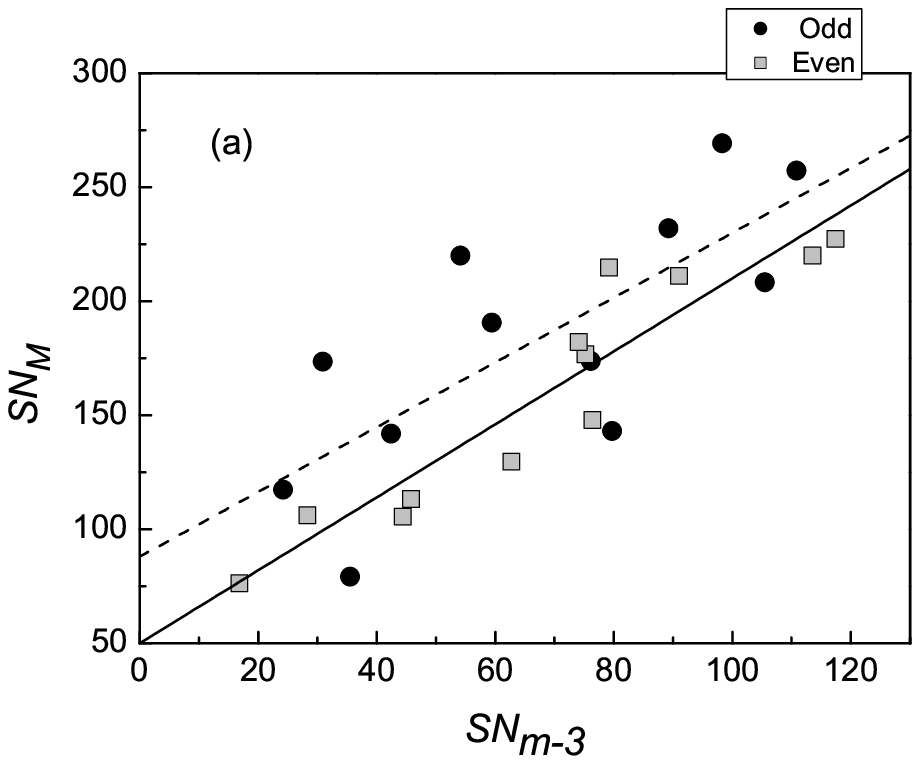}}
\fb{\includegraphics[width=0.45\textwidth]{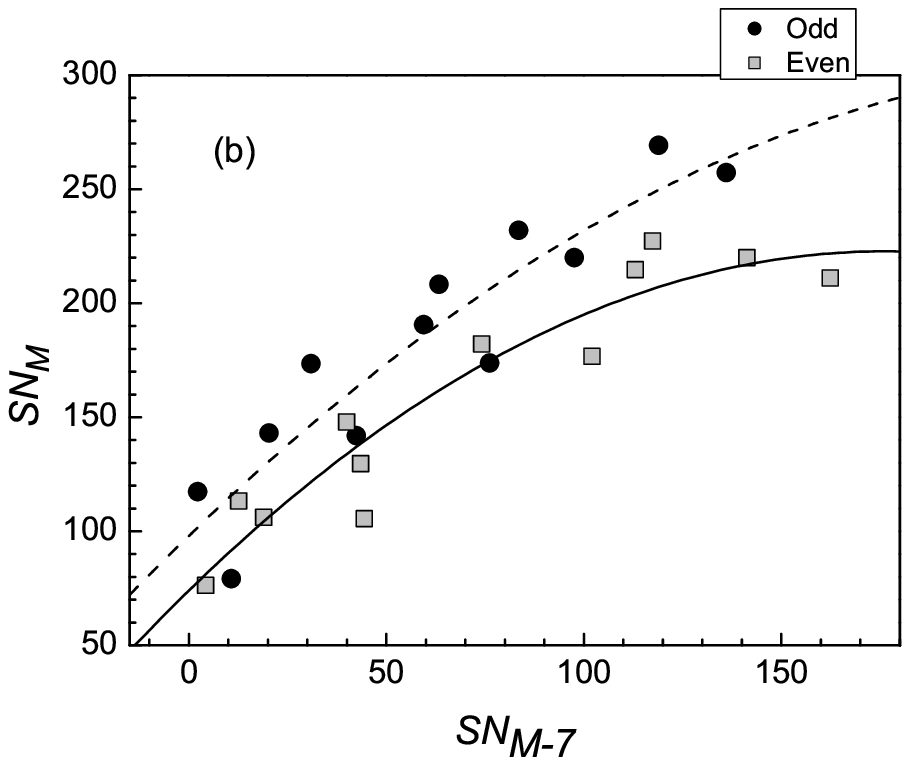}}
\end{center}
\caption{
(a) Dependence of $\SN_M$ on $\SN_{m-3}$ for even (solid line, gray squares)
and odd (dashed line, black circles) cycles. (b) Dependence of $\SN_{M-7}$ on
$\SN$ for even and odd cycles.
}
\label{fi3}
\end{figure}

This implies that \inlinecite{brajsa22}, having considered even and odd
cycles together, underestimated the predicted value of Cycle 25. The
difference in regressions for even and odd cycles is shown in
Figure~\ref{fi3}a.
 
Summarizing, according to the effect proposed by \inlinecite{brajsa22},
we can successfully predict the maximum of the next EVEN cycle three
years before the minimum, thereby supplementing the Gnevyshev-Ohl rule
with the relation even -- subsequent odd cycles. The success of the
prediction of an even cycle is guaranteed by a high correlation
coefficient of the equation~(\ref{eq3}).

The precursor of the prediction discussed in this section will be
called MI3E for even cycles and MI3O for odd ones (bearing in mind
that only MI3E corresponds to a reliable prediction).

\section{Cycle precursor associated with the maximum phase}

\begin{figure}
\begin{center}
\fb{\includegraphics[width=0.45\textwidth]{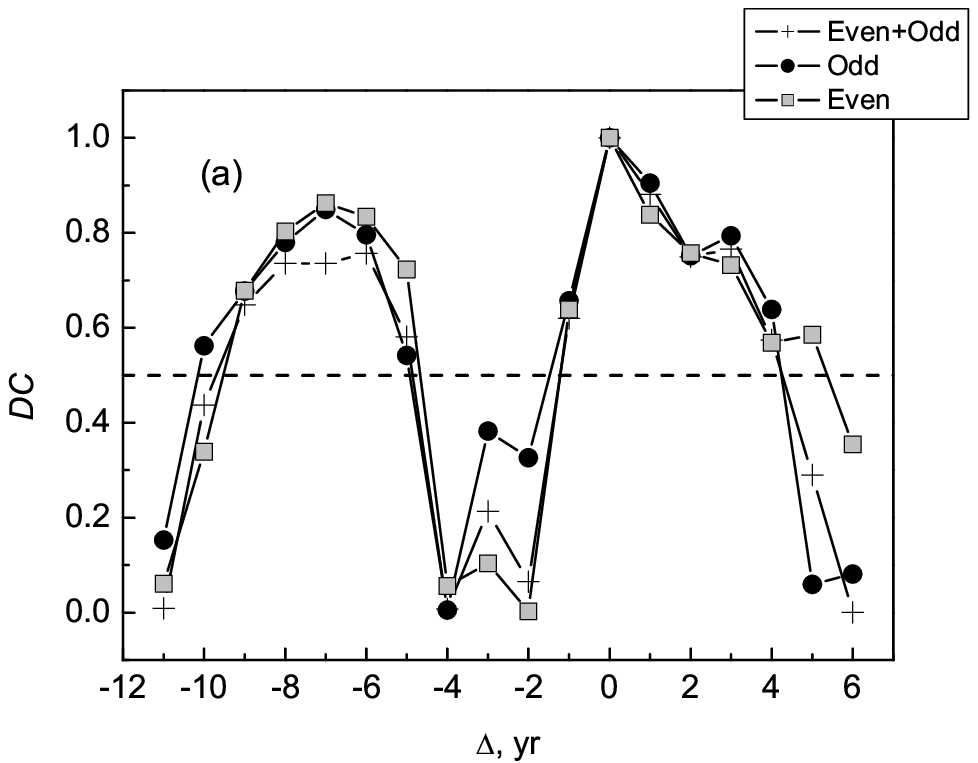}}
\fb{\includegraphics[width=0.45\textwidth]{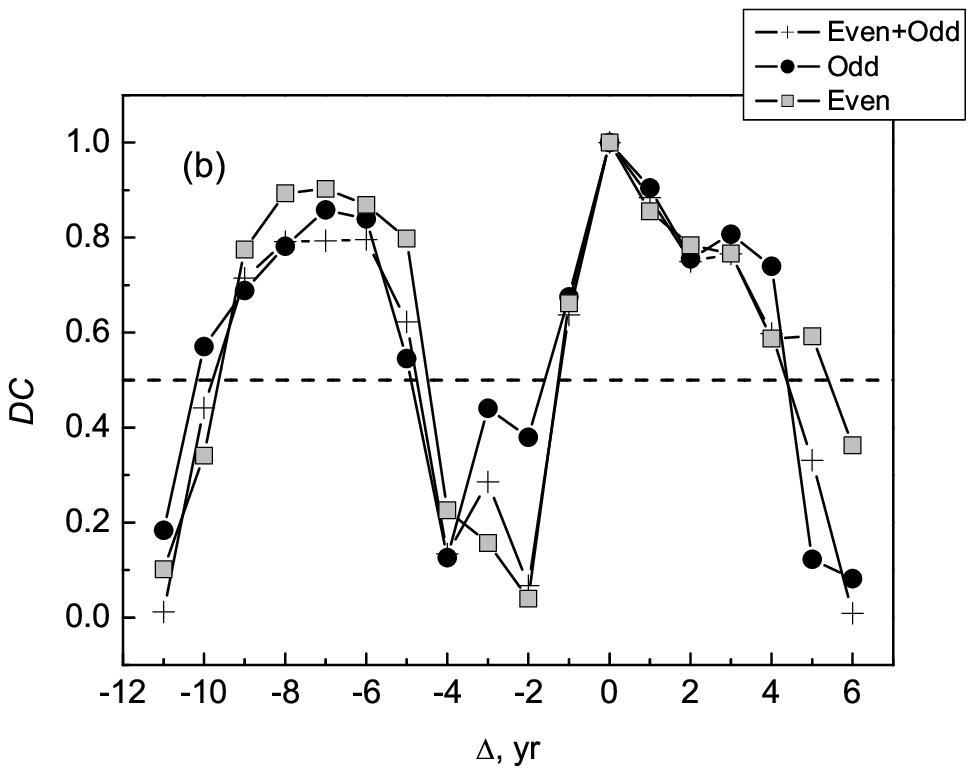}}
\end{center}
\caption{
Coefficients of the determination for relations $\SN_M$ with $\SN_{M+\Delta}$ for (a) linear form~(\ref{eq6}), (b) nonlinear forms~(\ref{eq7})--(\ref{eq8}).
}
\label{fi4}
\end{figure}

In the paper by \inlinecite{slonim84}, the idea was expressed to search for
cycle precursors a certain number of years before the maximum. Thus,
the possible predictors turn out to be associated with the maximum
phase, and not with the minimum phase, as in the previous section.
Here we will also consider regressions separately for even, odd cycles
and for all cycles together, choosing for analysis the same time
interval as before, since 1749. Figure~\ref{fi4}a illustrates the changes
in the coefficients of determination for linear relationships of the
type
\begin{equation}
\SN_M = a + b\cdot \SN_{M+\Delta}\,.
\label{eq6}
\end{equation}

Here we bear in mind that $\Delta = 0$ corresponds to the maximum of the
cycle. Precursors $\SN_{M-\Delta}$ correspond to negative $\Delta$ --- i.e., years before
maximum, positive ones indicate relations with post-maximum years,
which are also interesting for the prediction. In addition, we take
into account the weak nonlinearity of the relations. For negative $\Delta$ in
the form
\begin{equation}
\SN_M = a + b\cdot \SN_{M+\Delta} + c\cdot \SN_{M+\Delta}^2\,,
\label{eq7}
\end{equation}
and for positive ones --- 
\begin{equation}
\SN_{M+\Delta} = a + b\cdot \SN_M + c\cdot \SN_M^2\,.
\label{eq8}
\end{equation}

We do this specifically for the possible prediction of not only the
maximum values but also the index values in subsequent years, which,
as it turns out, are quite closely related to the maximum. The
determination coefficients for relations of type~(\ref{eq7})--(\ref{eq8}) are shown in
Figure~\ref{fi4}b.
 
Let's note two circumstances. The first --- in comparison with $\SN_{m-3}$ a more
``successful'' precursor for the prediction of the maximum is $\SN_{M-7}$: in the
case of a linear form~(\ref{eq6}) for odd cycles $DC=0.849$ ($\rho=0.921$), for even $DC=0.863$ ($\rho=0.929$); in the case of
a nonlinear form~(\ref{eq7}) for odd $DC=0.859$ ($\rho=0.927$), for even $DC=0.903$ ($\rho=0.950$). Secondly, when considering
all cycles together, the correlation of $\SN_{M-7}$ with $\SN_M$ is less than for each
of the parity separately, and this means that their relations are
different. This is clearly seen in Figure~\ref{fi3}b.

\begin{figure}
\begin{center}
\fb{\includegraphics[width=0.5\textwidth]{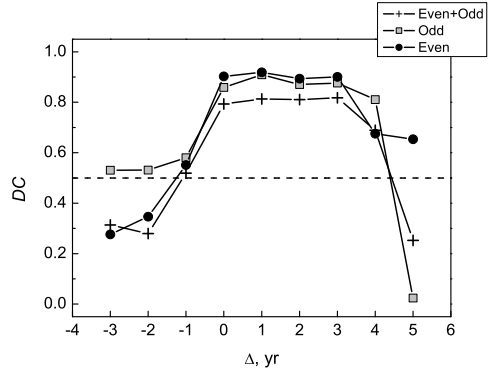}}
\end{center}
\caption{
Determination coefficients for quadratic relations of $\SN_{M-7}$ with $\SN_{M+\Delta}$.
}
\label{fi5}
\end{figure}

When we predict the amplitude of the future cycle, we are, of course,
primarily interested in the amplitude of the maximum, but we are also
interested in neighboring years. We found that the predictor of the
maximum is $\SN_{M-7}$. And how much can this value predict the years $M-1$,
$M+1$, $M+2$, etc.? Figure~\ref{fi5} answers this question: the seventh year
before the maximum is a predictor not only of the maximum but also of
the index in the years near the maximum. This applies to both even and
odd cycles, although the relations between $\SN_{M-7}$ and $\SN_{M+\Delta}$ are different.

The prediction precursor discussed in this section we will call MA7E for even cycles and MA7O for odd ones.

Thus, we have created a ``constructor'' for the prediction of the solar cycle, in this case --- the odd solar Cycle 25.

\section{Prediction of the solar activity cycle 25}

Let's use the results of the previous sections to predict Cycle 25. We
use the MA7O predictor. Assuming different years of maximum onset, we
get the $\SN_M$ values given in the Table~\ref{tab2}.

\begin{table}
\caption{
Predicted $\SN_M$ values depending on the estimated maximum year.
}\label{tab2}
\begin{tabular}{cc}     
\hline
Year & $\SN_M$ \\
\hline
2022.5 & $199 \pm 25$ \\
2023.5 & $159 \pm 25$ \\
2024.5 & $133 \pm 25$ \\
\hline
\end{tabular}
\end{table}

Then, using regressions, the coefficients of determination of which
are shown in Figure~\ref{fi5}, we calculate the values of $\SN$ in the years of
Cycle 25, which have prognostic significance, for different maximum
years from Table~\ref{tab2}.

\begin{figure}
\begin{center}
\fb{\includegraphics[width=0.45\textwidth]{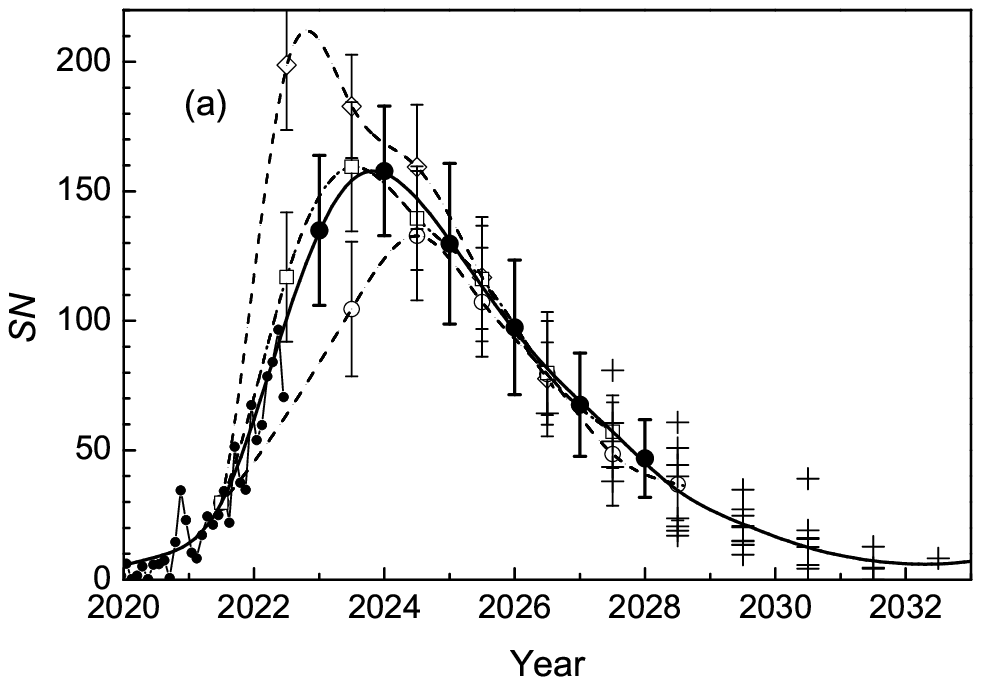}}
\fb{\includegraphics[width=0.45\textwidth]{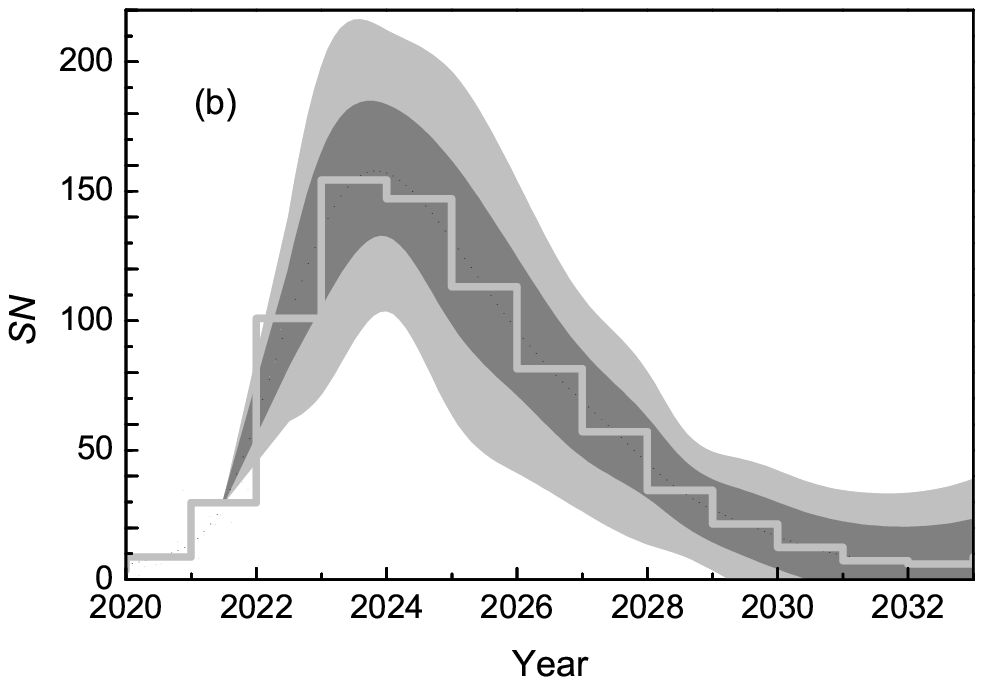}}
\end{center}
\caption{
(a) Various versions of predictions for the development of
Cycle 25 --- see the legend in the text; (b) The most probable
prediction of Cycle 25. Light gray areas are the prediction interval
with a confidence of 68\%, dark gray areas are the same with a
confidence of 95\%, and a stepped line is the most probable prediction
of average annual values.
}
\label{fi6}
\end{figure}

The results are shown in Figure~\ref{fi6}a: light icons --- rhombuses, squares,
and circles --- connected by a dashed line constructed using a global
cubic spline. Comparing the obtained curves with the course of the
monthly average values of $\SN$ observed up to June 2022 (small circles
connected by thin lines), we conclude that 2023.5 is the most
preferable of the three years of the assumed maximum from Table 2.
However, in general, the monthly average values are located, although
close to the interpolation curve, but still somewhat below her. Let's
assume a new maximum date --- 2024.0. Note that the average annual
values can be calculated by averaging $\SN$ not only traditionally from
January to December but also from July to June of the next year, and.
Let's calculate such annual averages, output regressions with
determination coefficients similar to those shown in Figure~\ref{fi5}, and
calculate the values for the dates 2023.0--2028.0 in annual
increments --- see Figure~\ref{fi6}a: black circles connected by a thick
spline curve. It can be seen that by assuming a maximum year of 2024.0
we get a better agreement of the behavior of the predicted and
observed $\SN$.

Now about the values indicated in Figure~\ref{fi6}a by crosses. Are there any
general trends in the last years of the cycle (in our case, the odd
one)? Consider the values of $\SN$ in the years of the final minima of
the odd cycle $m$ and the two years before it $m-1$ and $m-2$, depending
on their distance from the initial minimum $\Delta$. It turns out that this
dependence can be described by a parabola $\SN(\Delta) = (327\pm87)-(50\pm18)\Delta+(1.96\pm0.87)\Delta^2$
with a correlation coefficient $\rho=0.811$, i.e. sufficiently high, and we can use it to predict
the descending phase of a new cycle (as it seems, this is a new
unexpected result). In Figure~\ref{fi6}a crosses represent data given by the
date of the initial minimum of Cycle 25 as $2019.5\pm\Delta$. Note here that for even
cycles, a similar dependence also occurs, although the correlation
coefficient with the parabola is somewhat less --- $\rho=0.768$ . However, it can
also be used for solar cycle prediction in the future.

So far, for predictions, we have used confidence ranges of estimates
for a deviation equal to the standard deviation (i.e., with a
probability of falling into the interval equal to 68\%) --- as in \inlinecite{brajsa22},
as well as in many authors of solar cycle predictions.
In fact, if we want to achieve a reliability of at least 95\%, we need
to set a prediction interval based on at least two standard
deviations. Therefore, we will repeat all the previous procedures for
this new requirement. The average annual values of $\SN$, interpreted for
years with traditional calculation from January to December, with
confidence bands are shown in Figure~\ref{fi6}b.

Note that our methodology allows us to predict individual average
annual values of $\SN$ in a cycle, and not only the amplitude of its
maximum, as with most authors.

Having made a prediction of the Cycle 25, it is natural to check
whether the Gnevyshev-Ohl rule is fulfilled for a pair of Cycles
24--25. In Figure~\ref{fi1}a, the corresponding point is circled. As we can
see, our prediction does not contradict this well-known rule.

1
\section{Results}

The Gnevyshev-Ohl rule establishes a relation between adjacent 11-year
cycles in the 22-year ``magnetic'' Hale cycle as: ``an even cycle
determines the next odd one,'' and thus they form a pair. In this work,
we found that the behavior of an odd cycle determines the behavior of
the subsequent even one. Namely, 3 years before the minimum, the value
of $\SN$ in the odd cycle is associated with the value of the maximum in
the subsequent even cycle (correlation coefficient $\rho=0.94$. Similar to GOR,
in this sense, cycles are linked in pairs, but opposite to the Rule.
This is one of the important results of the work.

Based on this result, we propose to use $\SN_{m-3}$ (Figure~\ref{fi3}a) as a precursor of
the subsequent EVEN cycle during the descending phase of the odd one ---
we call this method MI3E. For the prediction of an odd cycle or a
prediction without parity (as in the article by \opencite{brajsa22}),
this method gives less reliable results.

To predict the ODD cycle, we propose to use the precursor of the
seventh year to its maximum MA7O --- $\SN_{M-7}$ during the descending phase of
the previous even cycle (Figure~\ref{fi3}b). It turned out that in this case,
we can predict the years near the maximum with a high correlation
coefficient ($\rho=0.90{-}0.94$). In addition, 7 years before the maximum, it is also
possible to predict an even cycle according to the MA7E precursor
(Figure~\ref{fi5}).

Also noteworthy is the similar behavior of $\SN$ for different cycles
near the final minimum of the cycle, depending on the distance from
the initial minimum. Thus, the proposed approaches make it possible to
predict cycles of different parity.

The current Cycle 25 in the year 2023 should reach a maximum of 154
average annual values with a prediction interval of 25 with a
confidence of 68\% and 53 with a confidence of 95\%. This is higher than
the original official prediction of NOAA/NASA/ISES but lower than the
updated prediction of \inlinecite{leamon21} according to the
methodology of \inlinecite{macint20}. In 2024, $\SN$ will be almost as
high as in 2023 --- 147 units, so with smaller time averaging scales,
the maximum will fall at the end of 2023. Here we note that the
proposed approach makes it possible to predict the values of $\SN$ in
individual years of the cycle, and not only the amplitude of its
maximum.

We have not discussed in this article the relationship between the
precursors $\SN_{m-3}$ and $\SN_{M-7}$, their relative position on the descending phase of
cycles, as well as their physical meaning --- this is the task of the
following studies.

Yury Nagovitsyn thanks the Ministry of Science and Higher Education of
the Russian Federation for financial support for project
075--15--2020--780.

\end{article}

\end{document}